\documentclass[prl,twocolumn,showpacs]{revtex4-1}
\usepackage{amsfonts,amsmath,mathrsfs,epsfig,amsbsy,bm,verbatim}

\newcommand{\ua}{\uparrow}
\newcommand{\da}{\downarrow}

\begin{document}

\title{Topological invariants for spin-orbit coupled superconductor nanowires}
\author{Sumanta Tewari$^{1}$}
\author{Jay D. Sau$^2$}

\affiliation{$^1$Department of Physics and Astronomy, Clemson University, Clemson, SC
29634\\
$^2$Department of Physics, Harvard University, Cambridge, MA 02138
}

\begin{abstract}
We show that a spin-orbit coupled semiconductor nanowire with Zeeman splitting and $s$-wave superconductivity is in symmetry class
BDI (and not D as is commonly thought) of the topological classification of band Hamiltonians. The class BDI allows for an integer $\mathbb{Z}$ topological invariant equal to the number of Majorana fermion (MF)
modes at \emph{each end} of the quantum wire protected by the chirality symmetry (reality of the Hamiltonian). Thus it is possible for this system
(and all other $d=1$ models related to it by symmetry) to have an arbitrary integer number, not just $0$ or $1$ as is commonly assumed, of MFs localized at each end of the wire. The integer counting the number of MFs at each end reduces to $0$ or $1$, and the class BDI reduces to D, in the presence of terms in the Hamiltonian that break the chirality symmetry.

\end{abstract}

\maketitle

\paragraph{Introduction:} Rashba spin-orbit (SO) coupled semiconductors in dimensions $d=2,1$ with a Zeeman field and proximity-induced $s$-wave superconductivity have recently attracted a lot of attention \cite{sau-et-al, Annals, Alicea-Tunable,long-PRB,roman,oreg,Lutchyn2,Alicea,Beenakker,Tewari-NJP,Mao,zhang-tewari,stanescu, Qu,Lutchyn3,Sau-Disorder,Heck}. Under
suitable external conditions these systems can support MF excitations (defined by second quantized operators $\gamma^{\dagger}=\gamma$)
whose statistics is non-Abelian. 
In $d=2$ the particle-hole (p-h) symmetric Bogoliubov-de Gennes (BdG) Hamiltonian of the Rashba-coupled semiconductor is
in the topological class D \cite{Schnyder,Kitaev} with an integer $\mathbb{Z}$ topological invariant which counts the number of
gapless chiral Majorana modes on the boundary. From dimensional reduction (i.e., by putting one of the wave-vectors to zero \cite{Ruy-Schnyder}), the gapless boundary Majorana modes in $d=2$ reduce to zero-energy \emph{end} Majorana modes in a $d=1$ nanowire.
The dimensional reduction argument therefore suggests that the number of possible end Majorana
modes in a SO coupled nanowire should also in principle be an integer. Based on this (and more rigorous arguments below) we argue that the Hamiltonian of the system in $d=1$ is in the topological
class BDI (in contrast to its being in class D in $d=2$), characterized by a $\mathbb{Z}$ topological invariant which counts the
number of possible end MFs in the nanowire.
By rigorous
arguments we clarify why the $d=1$ system is in class BDI and has a $Z$ topological invariant, construct an algebraic form of the
invariant, discuss its relation with the more-frequently-used $Z_2$ invariant, and illustrate these points by constructing explicit examples.
The work here significantly clarifies the topological properties of the SO coupled nanowires and related systems which have recently become
an important focus of attention both theoretically and experimentally.

The topological class of the SO coupled semiconductor is analogous to that of a spinless $p_x+ip_y$ superconductor. In $d=2$
the spinless $p_x+ip_y$ superconductor, with broken time reversal (TR) invariance (due to the presence of $i$ in the order parameter), is in class D characterized by a $\mathbb{Z}$ invariant
counting the number of gapless chiral Majorana modes on the boundary \cite{Volovik,Read}. It is also possible to define
a $\mathbb{Z}_2$ invariant which only counts the parity of the number of boundary Majorana modes \cite{Ghosh}.
Dimensional reduction arguments suggest that the number of possible end MFs in a $d=1$ spinless superconductor 
should also be an integer and this has recently been shown explicitly \cite{Sudip}.  Therefore,
the Hamiltonian should be in the topological
class BDI with a $\mathbb{Z}$ invariant in $d=1$.
  Note, however, that $d=1$ Hamiltonians in
class BDI are supposed to be TR-invariant while the spinless $p_x+ip_y$ superconductor explicitly breaks
TR symmetry in $d=2$. The key to this difference is that, in $d=1$, the Hamiltonian can be made completely real \cite{Sudip} while
it is necessarily complex in $d=2$. Redefining the time-reversal operator only in terms of the complex conjugation operator ${\cal K}$, it follows that in $d=2$ this symmetry is broken (class D) but it remains intact in $d=1$ (class BDI). More generally, the emergence of the reality condition in $d=1$ (reality of $H_{BdG}$ in the present case implies the chiral symmetry \cite{Schnyder,Kitaev,Ruy-Schnyder} given by ${\cal S}={\cal K} \cdot \Lambda$ where $\Lambda$ is the p-h transformation operator) changes the symmetry class of both the spinless $p$-wave superconductor and the Rashba-coupled system from D in $d=2$ to BDI in $d=1$.

Despite the fact that $H_{BdG}$ for a $d=1$ Rashba-coupled nanowire can be
made purely real, the $\mathbb{Z}$ invariant is not computable as a winding number of the Anderson pseudo-spin $\vec{d}$-vector defining the BdG Hamiltonian as $H_{BdG} (\mathbf{k})=\vec{d}(\mathbf{k}).\vec{\tau}$ \cite{Volovik,Anderson,Read}. Here, $\vec{\tau}$ is a vector of Pauli matrices defined in the p-h space.
This is because, in contrast to a spinless $p$-wave superconductor in $d=1$,
   the components of the $\vec{d}$-vector for the nanowire are themselves matrices. More generally, $H_{BdG}$ for a topological superconducting (TS) system in $d=1$ can be real (thus preserving the chiral symmetry) but can be a large $2N\times2N$ square matrix so the components of the $\vec{d}$-vector are themselves
   $N\times N$ matrices. 
   We show below how to compute the integer topological invariant for this problem in terms of a generalized pseudo-spin vector and its winding number and connect the value of this integer to the number of independent MF modes at each end of the nanowire. Note that multiple Majorana modes at \emph{each end} are protected by the chiral
    symmetry. We further show that the role of the usual $\mathbb{Z}_2$ Pfaffian invariant \cite{Kitaev-1D,Ghosh} reduces to determining the parity of the $\mathbb{Z}$ invariant.  In the presence of chiral symmetry breaking terms
    (i.e., terms which introduce complex entries in $H_{BdG}$) the topological class of the system reduces to D even in $d=1$ and an odd (even) number of MFs at each end becomes equivalent to just one (zero).

   \paragraph{$\mathbb{Z}$ invariant for real BdG Hamiltonians in $d=1$:}

To understand the difference between complex and real Hamiltonians let us start from the Hamiltonian of a spinless $p_x+ip_y$ superconductor
in $d=2$,
\begin{equation}
H_{1} (\mathbf{k})=(\epsilon_{\mathbf{k}}-\mu)\tau_z + \Delta_xk_x\tau_x -\Delta_yk_y\tau_y
\label{eq:H2DP},
\end{equation}
where $\mathbf{k}$ is a two-dimensional wave-vector, $\mu$ is the chemical potential, and $\Delta_x,\Delta_y$ are superconducting pair potentials along the $x,y$ directions, respectively. Here we have used the p-h basis $(c_{\mathbf{k}}^{\dagger},c_{-\mathbf{k}})$ and its
 hermitian conjugate, and the $\tau$ matrices in Eq.~\ref{eq:H2DP} are defined in this basis. Writing this Hamiltonian in terms of the Anderson
pseudo-spin vector \cite{Anderson} $\vec{d}(\mathbf{k})$ as  $H_{1} (\mathbf{k})=\vec{d}(\mathbf{k}).\vec{\tau}$, we see that for spinless $p_x+ip_y$ superconductor in $d=2$ all three components of $\vec{d}$ are non-zero.
The group of topological invariant is then $\mathbb{Z}$ which is the relevant homotopy group $\pi_2(S^2)$ of the mapping from the two-dimensional
  $k$ space to the 2-sphere of the 3-component unit vector $\hat{d}=\vec{d}/|\vec{d}|$ \cite{Volovik,Read}. On the other hand, in $d=1$, since the corresponding Hamiltonian can be made purely real ($\Delta_x$ drops out from Eq.~(\ref{eq:H2DP}) for the system along the $y$-axis), the vector $\vec{d}$ has only two components. Noting that the $k$-space now is also one-dimensional, the
    topological invariant must again be in $\mathbb{Z}$ (class BDI) since $\pi_1(S^1)=\mathbb{Z}$. This invariant is simply the winding number,
    \begin{equation}
    N=\frac{1}{2\pi}\int_0^{2\pi} d\theta(k),
    \label{eq:winding1}
    \end{equation}
    where $\theta(k)$ is the angle the unit vector $\hat{d}$ makes with, say, the $z$-axis on the $y-z$ plane.
    The winding number counts the number of times the 2-component vector $\hat{d}$ makes a complete cycle in its plane as $k$ varies in the one-dimensional
    Brillouin zone.
 It is clear that only with the breakdown of the reality condition of the BdG Hamiltonian (i.e., chiral symmetry given by ${\cal S}={\cal K} \cdot \Lambda$) the symmetry class of the spinless $p$-wave superconductor can change from BDI to D (which is characterized by a $\mathbb{Z}_2$ invariant) even in $d=1$.

 In $d=2,1$ the $4\times4$ BdG Hamiltonian $H_2(\mathbf{k})$ of a Rashba-coupled semiconductor with Zeeman coupling and proximity induced $s$-wave superconductivity is given by,
  \begin{eqnarray}  \label{eq:polar_bulk_H}
H_2(\mathbf{k})&=&(\epsilon_{\mathbf{k}}-\mu)\tau_z + V_Z \hat{\bm{S}}\cdot \bm\sigma \tau_z+\alpha k_x\sigma_y\tau_z
\nonumber\\&-&\alpha k_y\sigma_x+\Delta_0\sigma_y\tau_y,
\end{eqnarray}
where we have used the $4$-component p-h spinor $(u_\ua(\bm r),u_\da(\bm r),v_\ua(\bm r),v_\da(\bm r))$ (with
quasiparticle operators given by $d^{\dagger}=\sum_{\sigma}(u_{\sigma}(r)c_{\sigma}^{\dagger}(r)+v_{\sigma}c_{\sigma}(r))$),
and the Pauli matrices  $\sigma_{x,y,z},\tau_{x,y,z}$ act on the spin and particle-hole spaces, respectively.
In Eq.~(\ref{eq:polar_bulk_H}), the vector $\hat{\bm{S}}$ is a suitably chosen direction of the applied Zeeman spin splitting $V_Z$
(e.g., $\hat{\bm{S}}=\hat{z}$ in $d=2$ for $\mathbf{k}=(k_x,k_y)$, $\hat{\bm{S}}=\hat{x}$ in $d=1$ for $\mathbf{k}=k_x$),
$\mu$ is the chemical potential, $\alpha$ is the
Rashba SO coupling constant, and $\Delta_0$ is an $s$-wave superconducting pair-potential.
  It is clear that in $d=2$ it is not possible to make the Hamiltonian real because of the complex Rashba term. In contrast, in $d=1$ $H_{2}$ can be made purely real (Rashba term can be made real), and one can define a pseudo TR operator in terms of ${\cal K}$ alone. Then, in $d=1$ $H_{2}$ preserves both p-h as well as the new `time reversal' symmetry and hence is in class BDI characterized by a $\mathbb{Z}$ invariant. Note, however, that in contrast to the case of a spinless $p$-wave superconductor,
   the components of the $\vec{d}$-vector in the present $4\times4$ Hamiltonian are themselves $2\times2$ matrices. It is not clear how to define the winding number of the $\vec{d}$ vector or the $\mathbb{Z}$ invariant for such BdG Hamiltonians which are larger than simple $2\times2$ matrices as in spinless
   $p$-wave superconductor. More generally, the BdG Hamiltonian of a TS system in $d=1$, despite being real (thus preserving the chiral symmetry), can be a large $2N\times2N$ square matrix so that the components of the $\vec{d}$-vector are
   $N\times N$ matrices. We now show below how to compute the integer topological invariant for this problem by generalizing the concept of the $\vec{d}$-vector winding number for arbitrary dimensional matrices.

    For real BdG Hamiltonians in $d=1$ it is possible to construct an integer topological invariant
in terms of a generalized winding number. To see this, consider an arbitrary second quantized BdG Hamiltonian (real or complex) which can be written as \cite{Stone-Chung},
\begin{equation}
H=\left(\begin{array}{cc}H_{0,ij}&\Delta_{ij}\\\Delta^{\dagger}_{ij} & -H^{T}_{0,ij}\end{array}\right)
\label{eq:H1}
\end{equation}
To write the BdG matrix in Eq.~(\ref{eq:H1}) we have used the p-h basis $\Psi^{\dagger}_i,\Psi_j$ where $\Psi^{\dagger}_i=(c^{\dagger}_i, c_i)$,
$\Psi_j=(c_j,c^{\dagger}_j)^{T}$, $i,j$ are collective coordinates representing both spatial coordinates $r,r^{\prime}$ and spin indices $\uparrow,\downarrow$, and summations over $i$ and $j$ are implied. In this form, the gap matrix $\Delta_{ij}$ is necessarily an anti-symmetric matrix. If the BdG matrix is purely real, $H^{T}_{0,ij}=H_{0,ij}$ and $\Delta^{\dagger}_{ij}=\Delta^{T}_{ij}=-\Delta_{ij}$. In this case, we can
rewrite the BdG Hamiltonian in the form (omitting the $i,j$ indices)
\begin{equation}
H=\left(\begin{array}{cc}H_0&\Delta\\-\Delta & -H_0\end{array}\right)
\label{eq:HMatrix2}
\end{equation}
where $H_0$ is real-symmetric and $\Delta$ is real anti-symmetric. Such a Hamiltonian $H$ has the chiral symmetry \cite{Schnyder,Kitaev,Ruy-Schnyder} defined as $S={\cal K}\cdot\Lambda$ (with $\Lambda=\tau_x \cdot{\cal K}$ in this basis), under which the Hamiltonian is invariant.
  Since in the p-h space the matrix $H$ can be written as $H=H_0\tau_z + i\Delta\tau_y$ it can be made purely off-diagonal by a rotation in the p-h space by the unitary
transformation $U=e^{- i\frac{\pi}{4}\tau_y }$. It follows that the rotated Hamiltonian
\begin{equation}
U H U^\dagger = \left(\begin{array}{cc}0&A \\ A^T & 0\end{array}\right)
\end{equation}
is a symmetric Hamiltonian with the matrix $A=H_0+\Delta$ being real.

Fourier transforming to momentum space $A(k)$ satisfies the constraints $A(k)=A^*(-k)$, so that $A(K)$ is  real at the p-h symmetric points $k=K$, which are given by $K=0, \pm \pi$ in $d=1$.
The Hamiltonian in the k-space can be written as,
\begin{equation}
U H(k) U^\dagger = \left(\begin{array}{cc}0&A(k)\\ A^T(-k) & 0\end{array}\right).
\end{equation}
 Notice that the existence of a zero eigenvalue of the matrix $A(k)$ necessarily implies the existence of a zero
 eigenvalue of the BdG Hamiltonian. Therefore, as long as all the eigenvalues of $A(k)$ are gapped (which implies that the
 determinant of $A(k)$ is gapped), the Hamiltonian is also gapped. It follows that a non-zero $Det(A(k))$ for \emph{all} values of $k$ in the
 $d=1$ Brillouin zone indicates a fully gapped Hamiltonian.

 Since $A(k)$ is a purely real matrix at the p-h invariant points $k=0,\pm\pi$, $Det(A(k))$, which is in general a complex number for a general value of $k$, is also purely real for $k=0,\pm\pi$. We now write $Det(A(k))$ as $Det(A(k))=|Det(A(k))|\exp(i\theta(k))$. Note that $\theta(k)$ must be $0$
 or $\pm\pi$ at $k=0,\pm\pi$. Defining the variable
$
z(k)=\exp(i\theta(k))=Det(A(k))/|Det(A(k))|,
$
 we can now write the expression for a winding number,
\begin{equation}
W=\frac{-i}{\pi}\int_{k=0}^{k=\pi} \frac{d z(k)}{z(k)},\label{eq:W}
\end{equation}
which can only be an integer (i.e. $W\in Z$) including zero. Similar topological invariants for BDI Hamiltonians in $d=1$ have
 also been constructed earlier \cite{Zak,Hatsugai} This integer is a topological invariant because it is not related to any symmetry breaking and yet can only change when $Det(A(k))$ goes through zero somewhere between $k=0$ and $k=\pi$ which indicates a closure of the spectral gap and the associated quantum phase transition.

We now do a consistency check on the topological invariant $W$.
By straightforward algebra we can show that,
\begin{align}
&W=\frac{-i}{\pi}\int_0^\pi\frac{d z(k)}{z(k)}
=\int_0^\pi \frac{dk}{\pi}\partial_k [\textrm{log}z(k)]\nonumber\\
&=\int_0^\pi \frac{dk}{2\pi}Tr\Big[\tau_z \left(\begin{array}{cc}A(k)\partial_k A^{-1}(k)&0\\0& A^T(-k)\partial_k A^{T,-1}(-k) \end{array}\right)\Big]\nonumber\\
&=\int_0^\pi \frac{dk}{2\pi}Tr[\tau_z H(k)\partial_k H(k)^{-1}]
=W_1,
\end{align}
where $W_1$ is the topological invariant written for chiral systems in terms of the zero-frequency single-particle Green's functions $G(k)=H^{-1}(k)$
in Ref.~\onlinecite{Gurarie}. It can also be shown \cite{Gurarie} that the difference of the integers $W_1$ between two gapped topological systems gives the number of zero energy modes at a boundary separating them. It follows that $W$ gives the number of MF modes at each end of a semiconductor nanowire since the end separates the chiral nanowire (with integer invariant $W$) from vacuum (with $W=0$).

\paragraph{Pfaffian $\mathbb{Z}_2$ invariant as parity of the $\mathbb{Z}$ invariant:}

Now we derive a formula connecting the $\mathbb{Z}$ invariant $W$ and the Pfaffian
$\mathbb{Z}_2$ invariant \cite{Kitaev-1D,Ghosh} more frequently used for a semiconductor nanowire. For this, consider
 a BdG matrix $H_{BdG}$ with a particle-hole symmetry of the form $\tau_x H_{BdG}=-H_{BdG}^*\tau_x$ (note that $\Lambda={\cal K}\cdot \tau_x$) where $\tau_x=\tau_x^T$ is the symmetric
 particle-hole transformation matrix satisfying $\tau_x\tau_x^*=1$.
Then the matrix $H_{BdG}\tau_x$ is anti-symmetric i.e
\begin{equation}
(H_{BdG}\tau_x)^T=\tau_x H_{BdG}^*=-H_{BdG}\tau_x.
\end{equation}
This allows us to define a Pfaffian $Pf(H_{BdG}\tau_x)$ associated with the BdG Hamiltonian.

Now note that
\begin{align}
&Pf[H\tau_x]=Pf[U^\dagger \left(\begin{array}{cc}0& A(k)\\ A^T(-k) & 0\end{array}\right)U\tau_x U^T U^{*}]\nonumber\\
&= Pf[ \left(\begin{array}{cc}0&  A(k)\\ - A^T(-k) & 0\end{array}\right)]=Det(A(k))
\end{align}
at $k=0,\pi$, where we have used the fact that $Det(U)=1$. Then,
the Pfaffian topological invariant of the BdG Hamiltonian \cite{Ghosh}, which is the algebraic sign of the quantity $Q$,
\begin{equation}
Q=\frac{Pf[ \left(\begin{array}{cc}0&  A(k=0)\\ - A^T(k=0) & 0\end{array}\right)]}{Pf[ \left(\begin{array}{cc}0&   A(k=\pi)\\- A^T(k=\pi) & 0\end{array}\right)]},
\label{eq:Pfaffian}
\end{equation}
 is clearly equal to the sign of the $Det(A(k=0))/Det(A(k=\pi))$.
The sign of $Det(A(k=0))/Det(A(k=\pi))$ is in turn equal to the parity of $W$,
because, from Eq.~\ref{eq:W},
\begin{equation}
\textrm{sign}[\frac{Det(A(k=\pi))}{Det(A(k=0))}]=\frac{z(k=\pi)}{z(k=0)}=e^{i\pi W}=(-1)^W.
\end{equation}
It follows that the familiar $\mathbb{Z}_2$ Pfaffian invariant of the $d=1$ systems is simply the parity of the more
 general $\mathbb{Z}$ invariant
of a chiral Hamiltonian.

\paragraph{$\mathbb{Z}$ invariant for spinless superconductors in $d=1$:}
Next we show that $W$ as defined in Eq.~(\ref{eq:W}) reduces to the invariant $N$ (Eq.~\ref{eq:winding1})
for a $d=1$ spinless $p$-wave superconductor. In this case, the BdG Hamiltonian
is given by, $H_{BdG}=(\epsilon_k - \mu)\tau_z + \Delta(k)\tau_y$ where $\Delta(k)=\Delta_0k$ with $\Delta_0$ a constant. The $\vec{d}$-vector
then has just two components, $d_y=\Delta(k), d_z=\epsilon_k - \mu$.
To evaluate $W$ for this
system, note that $A(k)$ for the Hamiltonian is given by $A(k)=\epsilon_k - \mu + i\Delta(k)$. It therefore follows that the winding number associated
with the phase of the quantity, $z(k)=Det(A(k))/|Det(A(k))|
=\exp(i\theta(k))$ is the same as the winding number $N$ of the phase of the unit vector $\hat{d}(k)$ on the $d=1$ Brillouin zone.
The invariant $W$
vanishes for $\mu<0$ and is equal to $1$ for $\mu >0$ \cite{}. 
  The point $\mu=0$ marks a topological quantum transition at which $W$ changes by $1$. Note, however, that it is also possible for $W$ to jump by $2$ or
any other integer (for example, when $\Delta(k) \sim \Delta_0\sin k + \Delta_1\sin 2k$ with no relative phase between $\Delta_0$ and $\Delta_1$ \cite{Sudip}) giving rise to a topological transition between two ground states with number of MFs at each end differing by more than one.

\paragraph{$\mathbb{Z}$ invariant for the SO coupled nanowire:}
We now consider the case of a SO coupled semiconductor nanowire with a Zeeman coupling and a proximity-induced $s$-wave superconductivity
\cite{long-PRB,roman,oreg}.
In this case, from Eq.~(\ref{eq:polar_bulk_H}) and Eq.~(\ref{eq:HMatrix2}) we have,
$H_0=(\epsilon_k-\mu)+\alpha f(k) \sigma_y+V_Z\sigma_x$ and
 $\Delta=i\Delta_0\sigma_y$ so that $A(k)=(\epsilon_k-\mu)+\alpha f(k) \sigma_y+V_Z\sigma_x+i\Delta_0\sigma_y$. Here we have generalized
 the SO coupling term to have a general wave-vector dependence with the constraint $f(k\rightarrow \pm\pi)\rightarrow 0$. We find that
\begin{equation}
Det(A(k))=(\epsilon_k-\mu)^2+\Delta_0^2-V_Z^2-\alpha^2 f^2(k)+2i\Delta_0\alpha f(k)
\end{equation}
has a non-trivial winding number whenever the Pfaffian of the Hamiltonian (at $k=0$) $(\epsilon_k-\mu)^2+\Delta_0^2-V_Z^2-\alpha^2 f^2(k)<0.$
To see why this is so consider a simple model for the semiconductor
bandstructure $\epsilon_{k=0}=0$ and $\epsilon_{k=\pi/a}\gg \mu, \Delta,V_Z$.
Similarly, because the Rashba SO coupling is a result of broken
 inversion symmetry $f(k)=0$ for $k=0,\pi$.
Therefore we assume that $f(k)$ increases from $0$ to a
maximum value and then decreases as $k$ changes from $0$ to $\pi$. In the
case when the Pfaffian changes sign in going from  $k=0$ to $k=\pi$,
we see that  $Det(A(k))$ changes from being a negative real number (i.e. with
argument $Arg(Det(A(k=0)))=\pi$) to a positive real number
(i.e. with
argument $Arg(Det(A(k=\pi)))=0$) taking a route in the complex plane
over the positive real axis (i.e. $Im(Det(A(k)))>0$). Thus, from an
inspection of this trajectory, the winding of
the phase of $Det(A(k))$ i.e. $Arg(Det(A(k)))$ must equal $\pi$ and the winding number $W=1$. 
Since the nanowire is in the class BDI, the number
of MFs \emph{at each end} (and the invariant $W$)
can be \emph{any} positive integer (i.e, not just $0$ or $1$). Similar to the $p$-wave case \cite{Sudip,Volovik,Read} we can
construct an example of such a Hamiltonian (with $W=2$) by including the next-nearest-neighbor terms in the
dispersion and the SO coupling. Below we discuss a more realistic situation
characterized by $2$ MFs at each end of a quantum wire ($W=2$) the stability of which is protected by the chirality symmetry. The commonly used 
$\mathbb{Z}_2$ invariant is trivial in this case and is not adequate to describe the topological properties of the nanowire. 

\paragraph{$\mathbb{Z}$ invariant for the two-channel SO coupled nanowire:}
 Consider a
a quasi-1D  semiconductor wire with two pairs of relevant confinement-induced bands (two channels). The BdG Hamiltonian for this system
proximity-coupled to an $s$-wave superconductor is given by the appropriate generalization of Eq.~\ref{eq:polar_bulk_H}, written as
  \begin{eqnarray}  \label{eq:polar_bulk_H2}
H_2(k)&=&(\epsilon_{\mathbf{k}}-(\mu+\delta\mu)-\delta\mu\rho_z)\tau_z + V_Z \sigma_x \tau_z+\alpha k\sigma_y\tau_z
\nonumber\\&+&(\Delta_0+\Delta_{12}\rho_x)\sigma_y\tau_y,
\end{eqnarray}
where $\rho_{x,y,z}$ represents Pauli matrices that account for the inter-band degree of freedom.
The confinement energy splitting between the two bands is given by $\delta \mu$, while the pairing between the bands is given by $\Delta_{12}$.
For this Hamiltonian, 
 $A(k)=(\epsilon_k-(\mu+\delta\mu)-\delta\mu\rho_z)+\alpha f(k) \sigma_y+V_Z\sigma_x+i(\Delta_0+\Delta_{12}\rho_x)\sigma_y$.
Let's first consider no inter-channel pairing potential,  $\Delta_{12}=0$. In this case, the matrix $A(k)$ commutes with $\rho_z$, so that $Arg(Det(A(k)))=Arg(Det(A_{\rho_z=+1}(k)))+Arg(Det(A_{\rho_z=-1}(k)))$,
where $A_{\rho_z=\rho}(k)$ are $2\times 2$ matrices similar to the single-channel case with the channel index fixed to $\rho_z=\rho=\pm 1$.
The winding number of $Det(A(k))$ is just the sum of the winding numbers for each of the two channels $\rho_z=\rho=\pm 1$,  which is given by
$\frac{1}{2}[\textrm{sgn}{V_z^2-\Delta_0^2-(\mu+\delta\mu+\delta\mu\rho_z)^2}+1]$. For $|V_Z|>\textrm{min}[\sqrt{\Delta_0^2+\mu^2},\sqrt{\Delta_0^2+(\mu-2\delta\mu)^2}]$ the total winding number can become $2$ so that there are two MFs at end of the wire. This occurs for Zeeman splittings larger than $2|\delta\mu|$, $V_Z>\mu,\mu-\delta\mu$.

We see above that a pair of MFs at each end is expected when the channels are decoupled
or as has been recently shown at non-generic phase-transition points.~\cite{roman_fisher} However, a non-zero value of the
topological invariant $W$ clearly shows that the pair of MFs at each end remain at exactly
zero energy even for non-zero inter-band pairing
$\Delta_{12}$ as long as $\Delta_{12}$ is not large enough to close the bulk superconducting gap.
Although intuitively clear, we have checked this numerically by computing the invariant using Eq.~\ref{eq:W}. The Majorana
nanowire in this case has two MFs at each end of the wire confirming the validity of the topological class BDI.

\paragraph{Conclusion:}
We show that the Hamiltonian of a SO coupled topological semiconductor nanowire is in the topological class BDI with an integer
$\mathbb{Z}$ topological invariant. The $\mathbb{Z}$ invariant counts the number of possible zero-energy MFs \emph{at each end} of the nanowire. We show that the familiar $Z_2$ invariant of this system only gives the parity of the integer invariant. In contrast to the prevalent view, we find that under appropriate conditions multiple MFs (equal to the $\mathbb{Z}$ invariant) at each end of the wire can all be at zero energy protected by the chirality (reality) symmetry of the Hamiltonian. 

 We thank S. Chakravarty for enlightening discussions. S.T. thanks DARPA and NSF for support. J.S. thanks the Harvard Quantum Optics Center for support.


\end{document}